\documentclass[aps,preprint,nofootinbib]{revtex4-1} 
 
\usepackage{amsmath,amssymb}    
\usepackage{graphicx}   
\usepackage{verbatim}   
\usepackage{color}      
\usepackage{subfigure}  
\usepackage{hyperref}   
\usepackage{here}       
\usepackage{bm}         

\begin{document}

\preprint{KOBE-COSMO-16-07}

\title{Primordial Gravitational Waves Induced by Magnetic Fields \\
in an Ekpyrotic Scenario}
\author{Asuka Ito and Jiro Soda}
\affiliation{Department of Physics, Kobe University, Kobe 657-8501, Japan}

\date{May 29, 2017}

\begin{abstract}
\noindent \hrulefill 
\begin{center} \normalsize{Abstract} \end{center}

Both inflationary and ekpyrotic scenarios can account for the origin of the large scale structure of the universe.
It is often said that detecting  primordial gravitational waves is the key to distinguish both scenarios. 
We show that this is not true if the gauge kinetic  function is present  in the ekpyrotic scenario. 
In fact,  primordial gravitational waves sourced by the gauge field can be produced in an ekpyrotic universe. 
We also study scalar fluctuations sourced by the gauge field and 
show that it is negligible compared to primordial gravitational waves. 
This comes from the fact that the fast roll condition holds in  ekpyrotic models. 

\noindent \hrulefill 
\end{abstract}

\maketitle


\section{Introduction}
Inflation has succeeded in solving several issues in big bang cosmology and explaining the temperature anisotropy of 
the cosmic microwave background radiation (CMB) and the large scale structure of the universe. 
However, it is known that bouncing universe models \cite{Brandenberger:2016vhg}
such as the ekpyrotic scenario \cite{Khoury:2001wf} based on 
superstring theory \cite{Horava:1996ma} can do the same 
job \cite{Ijjas:2015hcc}.\footnote{The 
pre-big bang scenario is also a kind of the models \cite{Gasperini:2002bn}.
Our conclusion could apply to it too.} 
Therefore, it is important to clarify which scenario is actually realized in the early stage of the universe.

In the ekpyrotic scenario, the primordial fluctuations are produced in a slowly contracting (ekpyrotic) phase. 
The spectrum of the scalar and tensor vacuum fluctuations become blue-tilted at the end of the ekpyrotic phase. 
We therefore need an additional scalar field to explain the temperature anisotropy of 
the CMB~\cite{Levy:2015awa}. 
In the ekpyrotic scenario, the amplitudes of primordial gravitational waves \cite{Starobinsky:1979ty}
are quite small
and practically unobservable \cite{Boyle:2003km}. Hence, it is often said that, 
if we could detect the primordial gravitational waves, we would be able to disprove the ekpyrotic scenario.  
However,  if there could exist another mechanism for producing  gravitational waves
in the ekpyrotic scenario, the story would be completely different.  
Indeed, we show that there exists a mechanism for producing abundant gravitational waves in the ekpyrotic phase. 

The key is the presence of magnetic fields in the early universe.
Observationally, there are several evidences for magnetic fields to exist on various cosmological scales \cite{Subramanian:2015lua}. 
Although the origin of primordial magnetic fields is unknown,
the  presence of magnetic fields on extra galactic scales \cite{Chen:2014rsa} 
implies that the seed of magnetic fields must be produced in the early universe. 
Notably, there are  attempts to make primordial magnetic fields with the gauge kinetic  function 
in an inflationary universe \cite{Ratra} or in a bouncing universe \cite{Membiela:2013cea}. 

In this paper, we first show that scale invariant magnetic fields 
can be produced in the ekpyrotic phase in the presence of  the gauge kinetic function. 
Next, we show that the magnetic fields can become a source of abundant gravitational waves 
(such  mechanism works also in inflation \cite{Ito:2016aai}). 
It turns out that the gravitational wave spectrum is nearly scale invariant (slightly blue) 
at the end of the ekpyrotic phase. 
Hence, it is difficult to discriminate between inflation and the ekpyrotic scenario by 
merely detecting   primordial gravitational waves. 
We also study  scalar fluctuations induced by the magnetic fields and
show that the sourced tensor to scalar ratio should be more than unity, which implies that
scalar fluctuations  in  the CMB should be dominated by quantum fluctuations
produced by an additional scalar field as is often assumed in the ekpyrotic scenario. 

The paper is organized as follows. 
In section II, we review the ekpyrotic scenario briefly and explain background evolution in the ekpyrotic phase. 
In section III, we derive the mode functions of the gauge field and 
show that scale invariant magnetic fields can be produced in the ekpyrotic scenario. 
In section IV, we demonstrate that abundant gravitational waves with scale invariance are produced from 
the scale invariant magnetic fields. 
In section V, we show that  scalar fluctuations are also produced by the scale invariant magnetic fields. 
It turns out that  the tensor to scalar ratio should be larger than unity in the ekpyrotic scenario. 
The final section is devoted to the conclusion.
\section{Ekpyrotic Phase} 
In the ekpyrotic scenario, two branes residing in an extra dimension approach, collide and bounce off to each other. 
From the four-dimensional point of view, they correspond to a contracting universe and an expanding universe,  respectively. 
The ekpyrotic scenario can be described by a four dimensional 
 effective theory with a scalar field $\phi$ moving in an effective potential  $V(\phi)$ specified below. 
The action reads 
\begin{equation}
    S=\int d^{4}x \sqrt{-g}\left[ 
                     \frac{M_{pl}^{2}}{2}R-\frac{1}{2}(\partial_{\mu}\phi)(\partial^{\mu}\phi)
                          - V(\phi)  \right] \ ,  \label{action}
\end{equation}
where $M_{pl}$ represents the reduced Planck mass, $g$ is the determinant of the metric $g_{\mu\nu}$ , and 
$R$ is the Ricci scalar. 
The scalar field represents the separation $l$ between two branes $l \sim e^{\phi}$. 
The contracting universe ($\dot{\phi}<0$) is connected to the expanding universe ($\dot{\phi}>0$) through a bounce 
(a collision of two branes). 
The scalar and tensor vacuum fluctuations are produced in the  contracting phase where 
the scalar field rolls down a negative steep potential 
\begin{equation}
    V(\phi) \simeq V_{0}e^{\lambda \frac{\phi}{M_{pl}}}\ , \label{Veq}
\end{equation}
where $V_0$ is a negative constant.
Note that $\lambda$ is also negative and satisfies the fast roll condition $|\lambda| \gg 1$ to keep isotropy of the universe. 
Thus, we can take an isotropic metric ansatz in this phase as 
\begin{equation}
    ds^{2}=a(\tau)\left[ -d\tau^{2}+dx^{2}+dy^{2}+dz^{2} \right] \  ,
\label{eq2}
\end{equation}
where we used a conformal time $\tau$.
It is straightforward to derive  scaling solutions from Eqs.(\ref{action}) $\sim$ (\ref{eq2})
\begin{equation}
    a(\tau)=a_{end} \left( \frac{-\tau}{-\tau_{end}} \right)^{\frac{2}{\lambda^{2}-2}}, \ \ \ 
             \frac{\phi(\tau)}{M_{pl}}=\phi_{0}-\frac{2\lambda}{\lambda^{2}-2} \ln(-M_{pl}\tau) \ , 
             \label{scasol}
\end{equation}
where $\tau_{end}\ (<0)$ and $a_{end}$ represent the moment and the scale factor 
at the end of the ekpyrotic phase, respectively. 
The obtained vacuum scalar and tensor power spectrums are blue-tilted, so that we need an additional 
scalar field to explain the CMB observation \cite{Levy:2015awa}. 
Then, the ekpyrotic scenario predicts the nearly scale invariant scalar power spectrum and 
the blue-tilted tensor power spectrum. 
The situation is different from inflation where both spectra are nearly scale invariant. 
\section{Scale Invariant Magnetic Fields}
In this section, we show that scale invariant magnetic fields can be produced from quantum fluctuations
due to interaction between a scalar field and a gauge field in the ekpyrotic phase. 
We consider  the action 
\begin{equation}
    S=\int d^{4}x \sqrt{-g}\left[ 
                     \frac{M_{pl}^{2}}{2}R-\frac{1}{2}(\partial_{\mu}\phi)(\partial^{\mu}\phi)
                          - V(\phi) -\frac{1}{4}  f^{2}(\phi) F_{\mu\nu}F^{\mu\nu}
                              \right] \ ,  \label{action0}
\end{equation}
where $F_{\mu\nu}=\partial_{\mu}A_{\nu}-\partial_{\nu}A_{\mu}$ is the field strength of  the gauge field coupled to the scalar field $\phi$ and 
$f(\phi)$ represents the gauge kinetic function. 
Now, let us take the gauge kinetic  function as exponential type functional form which is 
ubiquitous in models obtained from dimensional reduction 
\begin{equation}
    f(\phi)=f_{0}e^{\rho \frac{\phi}{M_{pl}}}\ , \label{cup}
\end{equation}
%
where it has been set to be unity at the end of the ekpyrotic phase, and then 
there is no strong coupling problem.
We treat the gauge field as a test field and ignore the back reaction from the gauge field in the background.
Thus,  using the background solution (\ref{scasol}), we can express
 the gauge kinetic function as 
\begin{equation}
    f(\phi)\propto (-\tau)^{-\frac{2\rho\lambda}{\lambda^{2}-2}}\ . \label{ftau}
\end{equation}
Let us  expand the gauge field in Fourier space as 
\begin{equation}
    \vec{A}(\tau,\vec{x})=\int \frac{d^{3}k}{(2\pi)^{3/2}} 
                          {\bm{A}}_{\bm{k}}(\tau) e^{i \bm{k} \cdot \bm{x}}  \ .
\end{equation}
Then the part for the gauge field in the action (\ref{action0}) can be rewritten as 
\begin{equation}
    S_{gauge}=\frac{1}{2} \int d\tau d^{3}k f^{2}(\phi) 
    \left[ \bm{A}'_{\bm{k}}\bm{A}'_{-\bm{k}}-
    k^{2}\bm{A}_{\bm{k}}\bm{A}_{\bm{-k}} \right] \ ,\label{fac}
\end{equation}
where a prime represents a derivative with respect to the conformal time. 
The Fourier mode of the gauge field can be promoted into the operator and
 expanded by the creation and annihilation operators satisfying  commutation relations
$\left[ \hat{a}^{(\sigma)}_{\bm{k}},\hat{a}^{(\rho)\dagger}_{\bm{k}'} \right]=
\delta_{\sigma\rho}\delta(\bm{k}-\bm{k}')$ as 
\begin{equation}
    \hat{\bm{A}}_{\bm{k}}(\tau)=\sum_{\sigma=+,-} \vec{e}^{(\sigma)}(\hat{k}) \left[ 
                                  U_{k}(\tau)\hat{a}^{(\sigma)}_{\bm{k}}+
                                  U_{k}^{*}(\tau)\hat{a}^{(\sigma)\dagger}_{-\bm{k}} \right]\ ,
\end{equation}
where $\sigma$ represents the two polarization degrees of freedom of the gauge field. 
The circular polarization vectors $\vec{e}^{\ (\sigma)}$  satisfy the relations
\begin{eqnarray}
    \vec{k}\cdot \vec{e}^{\ (\pm)}(\hat{k})&=& 0 \quad ,\nonumber \\
    \vec{k}\times \vec{e}^{\ (\pm)}(\hat{k})&=&\mp i k \vec{e}^{\ (\pm)}(\hat{k})\quad ,\nonumber \\
    \left(\vec{e}^{\ (\pm)}(\hat{k})\right)^{*}=\vec{e}^{\ (\pm)}(-\hat{k})\ ,&& \  
                \left(\vec{e}^{\ (\sigma)}(\hat{k})\right)^{*} \cdot \vec{e}^{\ (\rho)}(\hat{k})
                  =\delta_{\sigma\rho} \quad .
\end{eqnarray}
The mode functions obey the equations  derived from the action (\ref{fac})
\begin{equation}
    U''_{k}+2\frac{f'}{f}U'_{k}+k^{2}U_{k}=0 \label{modef} \ .
\end{equation}
Using  new variables $u_{k}\equiv f U_{k}$ , we get 
\begin{equation}
    u_{k}''+\left( k^{2}-\frac{f''}{f}\right) u_{k}=0 \label{mode}\ .
\end{equation}
Substituting Eq.(\ref{ftau}) into  Eq.(\ref{mode}) and solving it with the Bunch-Davies initial 
condition, we get the mode functions
\begin{equation}
    u_{k}(\tau)=\frac{1}{\sqrt{2k}}\sqrt{\frac{-k\tau \pi}{2}}
                    H_{-\frac{1}{2}\frac{\lambda^{2}+4\rho\lambda-2}{\lambda^{2}-2}}^{(1)}(-k\tau) \label{modesol}\ ,
\end{equation}
where $H_{\gamma}^{(1)}(x)$ is the Hankel function of the first kind. 
Now, we define the electric and magnetic fields  as
\begin{equation}
    \vec{E}(\tau, \bm{x})\equiv -\frac{f}{a^{2}}\partial_{\tau} \vec{A}(\tau,\bm{x})\ , \quad
    \vec{B}(\tau, \bm{x})\equiv \frac{f}{a^{2}}\left( \nabla\times \vec{A}(\tau,\bm{x}) \right) \ \label{dif}.
\end{equation}
They can be expanded in Fourier space as 
\begin{eqnarray}
    \vec{E}(\tau,\bm{x})&=&\int \frac{d^{3}k}{(2\pi)^{3/2}} \hat{\bm{E}}_{\bm{k}}(\tau)
                          e^{i\bm{k}\cdot\bm{x}} \ \label{E},\\
    \vec{B}(\tau,\bm{x})&=&\int \frac{d^{3}k}{(2\pi)^{3/2}} \hat{\bm{B}}_{\bm{k}}(\tau)
                          e^{i\bm{k}\cdot\bm{x}} \ ,\label{BB}
\end{eqnarray}
where 
\begin{eqnarray}
    \hat{\bm{E}}_{\bm{k}}(\tau)&=&\sum_{\sigma=+,-}\vec{e}^{(\sigma)}(\hat{k})
                             \left[ \mathcal{E}_{k}\hat{a}^{(\sigma)}_{\bm{k}}+
                              \mathcal{E}^{*}_{k}\hat{a}^{(\sigma)\dagger}_{-\bm{k}}\right] \ ,\\
    \hat{\bm{B}}_{\bm{k}}(\tau)&=&\sum_{\sigma=+,-}\sigma \vec{e}^{(\sigma)}(\hat{k})
                              \left[ \mathcal{B}_{k}\hat{a}^{(\sigma)}_{\bm{k}}+
                              \mathcal{B}^{*}_{k}\hat{a}^{(\sigma)\dagger}_{-\bm{k}}\right] \ . \label{B}
\end{eqnarray}
Here, we defined
\begin{eqnarray}
    \mathcal{E}_{k}(\tau)=-\frac{f}{a^{2}}\partial_{\tau}U_{k}(\tau) \ ,\quad 
    \mathcal{B}_{k}(\tau)=\frac{f}{a^{2}} k U_{k}(\tau) \ .
\end{eqnarray}
Using the mode functions (\ref{modesol}), we obtain 
\begin{eqnarray}
    \mathcal{E}_{k}(\tau)&=&-\frac{\sqrt{\pi}}{2} k^{1/2} (-k\tau)^{1/2} a_{end}^{-2} 
       \left( \frac{-\tau}{-\tau_{end}} \right)^{-\frac{4}{\lambda^{2}-2}} 
       H_{\frac{1}{2}\frac{\lambda^{2}-4\rho\lambda-2}{\lambda^{2}-2}}^{(1)}(-k\tau)\ , \\
    \mathcal{B}_{k}(\tau)&=&\frac{\sqrt{\pi}}{2} k^{1/2} (-k\tau)^{1/2} a_{end}^{-2} 
       \left( \frac{-\tau}{-\tau_{end}} \right)^{-\frac{4}{\lambda^{2}-2}} 
       H_{-\frac{1}{2}\frac{\lambda^{2}+4\rho\lambda-2}{\lambda^{2}-2}}^{(1)}(-k\tau)\ \ .\label{elemag}
\end{eqnarray}
In the superhorizon limit $|k\tau|\rightarrow 0$,  we can use an approximation
\begin{equation}
    H^{(1)}_{\gamma}(x)\simeq 
                             -\frac{i\Gamma(-\gamma)}{\pi}e^{-i\pi \gamma} 
                             \left( \frac{x}{2}\right)^{\gamma} 
                             \label{hankel}\ .
\end{equation}
Then the magnetic fields are given by 
\begin{eqnarray}
    \mathcal{B}_{k}(\tau)&=&-\frac{(\lambda^{2}-2)^{2}}{\sqrt{\pi}}i 
       2^{-\frac{1}{2}\frac{5\lambda^{2}-4\rho\lambda-10}{\lambda^{2}-2}}
       \Gamma\left(\frac{1}{2}\frac{\lambda^{2}+4\rho\lambda-2}{\lambda^{2}-2}\right)
       e^{\frac{\pi i}{2}\frac{\lambda^{2}+4\rho\lambda-2}{\lambda^{2}-2}} \nonumber \\
       &&\times\ 
       k^{\frac{1}{2}\frac{\lambda^{2}-4\rho\lambda-2}{\lambda^{2}-2}}
       (-\tau)^{-\frac{2(\rho\lambda+2)}{\lambda^{2}-2}}
       (-\tau_{end})^{\frac{2\lambda^{2}}{\lambda^{2}-2}}H_{end}^{2}\ \ ,\label{mag0}
\end{eqnarray}
where we used a relation 
\begin{equation}
    a=\left( \frac{2}{\lambda^{2}-2} \right)\frac{1}{\tau H}\ . \label{rela}
\end{equation}
In order to obtain the scale invariant magnetic fields, we require 
\begin{equation}
    \rho=\frac{\lambda^{2}-2}{\lambda}\ \label{rho}.
\end{equation}
In this case, the electric fields are always subdominant compared with the magnetic fields. 
So we only consider the magnetic fields as the source of gravitational waves.
Substituting Eq.(\ref{rho}) into Eq.(\ref{mag0}), we get 
\begin{equation}
    \mathcal{B}_{k}(\tau)=\frac{3\sqrt{2}}{8}(\lambda^{2}-2)^{2} k^{-3/2} 
    \left( \frac{-\tau}{-\tau_{end}} \right)^{-\frac{2\lambda^{2}}{\lambda^{2}-2}} H_{end}^{2}\ \ . \label{mag}
\end{equation}
We find that the steeper the potential becomes, namely $|\lambda |$ is bigger, 
the more magnetic field is amplified. 

To avoid destroying the background evolution by back reaction from the magnetic fields, at least, 
we need the condition that the energy density of the electromagnetic field does not exceed that of the scalar field at the 
end of the ekpyrotic phase
\begin{eqnarray}
    <\rho_{em}>&=&\frac{1}{2}\frac{1}{(2\pi)^{3}}\int^{k_{end}}_{k_{in}}2\times \mathcal{B}_{k}(\tau_{end})^{2} d^{3}k 
                          \nonumber\\
             &=&\frac{9}{64\pi^{2}}(\lambda^{2}-2)^{4}H_{end}^{4} \ln\left[\frac{k_{end}}{k_{in}}\right] \nonumber \\
             &<& 3M_{pl}^{2}H_{end}^{2} \quad , \label{backreac}
\end{eqnarray}
where $k_{in}$ and $k_{end}$ represent the scales where a mode exits the Hubble horizon at the beginning and at the end of 
the ekpyrotic phase, respectively. 
%
%
%
Let us also check the back reaction problem in the equation for the scalar field.
The hamiltonian constraint and the equation for the scalar field are given as 
\begin{eqnarray}
  H^{2} &=& \frac{1}{3M_{pl}^{2}}\left( \frac{1}{2}\dot{\phi}^{2}+
  V(\phi) + <\rho_{em}> \right) \ , \label{hamihami} \\
  \ddot{\phi} &=& -3H\dot{\phi} - V_{,\phi} +\frac{2\rho}{M_{pl}}<\rho_{em}> \ ,  \label{scaeq}
\end{eqnarray}
where an overdot denotes a derivative with respect to the cosmic time.
Asuuming that the gauge field is negligible in Eq.(\ref{hamihami}), 
the fast roll condition let Eq.(\ref{hamihami}) be 
\begin{equation}
   \frac{1}{2} \dot{\phi}^{2} \simeq -V(\phi) \ . \label{fafa}
\end{equation}
Differentiating the both sides of Eq.(\ref{fafa}) with respect to the time, we find that
the first term is negligible compared with the second term in the right-hand side of Eq.(\ref{scaeq}).
Thus, we have the relation
\begin{equation}
  <\rho_{em}> \simeq \frac{M_{pl}}{2\rho}V_{,\phi} \ ,
\end{equation}
when the gauge field is significant in Eq.(\ref{scaeq}).
Then the ratio of the energy density of the gauge field to that of the scalar field is 
\begin{equation}
  \frac{<\rho_{em}>}{3M_{pl}^{2}H^{2}}\  \gg \   \frac{<\rho_{em}>}{V} 
        \simeq \frac{M_{pl}}{2\rho}\frac{V_{,\phi}}{V} 
        = \frac{1}{2}\frac{\lambda}{\rho} \ .
\end{equation}
The most right term is order unity in our scenario since $\lambda$ and $\rho$ are same order from Eq.(\ref{rho}).
Therefore, as far as the ratio of the energy density of the gauge field to that of the scalar field is small, 
the gauge field can be treated as a test field in any equations.
%
%
%

Taking a look at Eq.(\ref{backreac}),
for example, if we set $H_{end}=10^{-5}M_{pl}$, we obtain the minimum value of $\lambda$ about $-17$. 
Then the amplitude of the magnetic field at the end of the ekpyrotic phase is about $10^{49}$ G.
Thus,  the cosmological magnetic fields observed at present can be produced in the ekpyrotic scenario
\cite{Subramanian:2015lua}. 
Remarkably, such magnetic fields  can also induce abundant primordial gravitational waves. 
We will see it in the next section. 
\section{Gravitational Waves from Magnetic Fields}
In this section, we calculate the gravitational waves induced by the magnetic fields
 studied in the previous section. 
The method is similar to that used in inflationary universe \cite{Ito:2016aai,barnaby}.
One can get the tensor sector of the action (\ref{action0}) as 
\begin{equation}
    S_{GW}=\int d\tau d^3x \left[\  \frac{M^{2}_{pl}}{8}a^{2} 
       \left( h'_{ij}h'^{ij}-\partial_{k} h_{ij} \partial_{k} h^{ij} \right)   
            +\frac{1}{2}a^4\left( E_{i}E_{j}+B_{i}B_{j} \right) h^{ij} \ \right] \ \label{action5},
\end{equation}
where $h_{ij}$ is the  transverse traceless tensor and we used the definition of the 
electric and magnetic fields (\ref{dif}). 
The tensor fluctuation can be expanded in Fourier space as 
\begin{eqnarray}
    h_{ij}&=&\sum_{\sigma=+,-} \int \frac{d^3k}{(2\pi)^{3/2}} 
           \hat{h}_{\bm{k}}^{(\sigma)} e^{i \bm{k}\bm{x}} \Pi ^{(\sigma)}_{ij} \ , \label{hfourier}\\
    \hat{h}_{\bm{k}}^{(\sigma)}(\tau)&=&V_{k}(\tau)\hat{a}_{\bm{k}}^{(\sigma)}+
                                  V_{k}^{*}(\tau)\hat{a}_{-\bm{k}}^{(\sigma) \dagger}\ ,
\end{eqnarray}
where $\Pi ^{(\sigma)}_{ij}$ 
are polarization tensors constructed by circular polarization vectors as 
$\Pi ^{(\sigma)}_{ij}\equiv e_{i}^{(\sigma)}e_{j}^{(\sigma)}$ and 
we have used creation and annihilation operators. 
Substituting  Eqs.(\ref{E})$\sim$(\ref{B}) and (\ref{hfourier}) into  Eq.(\ref{action5}), we obtain 
\begin{eqnarray}
    S_{GW}=\sum_{\sigma=+,-}&&\int d\tau d^3k \biggl[\ \frac{M^{2}_{pl}}{4}a^{2} 
    \left( \hat{h}'^{(\sigma)}_{\bm{k}}\hat{h}'^{(\sigma)}_{\bm{-k}}-
    k^2 \hat{h}^{(\sigma)}_{\bm{k}}\hat{h}^{(\sigma)}_{-\bm{k}} \right)  \nonumber \\
   -  \frac{a^4}{2}&&\int \frac{d^{3}p}{(2\pi)^{3/2}}
    \left( \hat{E}_{i,\bm{p}}\hat{E}_{j,\bm{k}-\bm{p}}+\hat{B}_{i,\bm{p}}\hat{B}_{j,\bm{k}-\bm{p}} \right)
     e^{*(\sigma)}_{i}(\hat{k})e^{*(\sigma)}_{j}(\hat{k}) 
     \ \hat{h}_{\bm{-k}}^{(\sigma)} \ \biggr] \ .
\end{eqnarray}
Using the variable $v_{k}\equiv \frac{M_{pl}}{2}aV_{k}$\ , 
we can get the equation for the mode function of the gravitational waves as 
\begin{equation}
    v_{k}''(\tau)+\left( k^{2}-\frac{a''}{a} \right) v_{k}(\tau) 
                   = S^{(\sigma)}(\tau,\bm{k}) \label{hmode}\ ,
\end{equation}
where the source term is defined by 
\begin{equation}
    S^{(\sigma)}(\tau,\bm{k}) = -\frac{a^{3}}{M_{pl}} \int \frac{d^{3}p}{(2\pi)^{3/2}} 
                                    \left( \hat{E}_{i,\bm{p}}\hat{E}_{j,\bm{k}-\bm{p}}+\hat{B}_{i,\bm{p}}\hat{B}_{j,\bm{k}-\bm{p}}\right)
                            e^{*(\sigma)}_{i}(\hat{k})e^{*(\sigma)}_{j}(\hat{k}) \ .
\end{equation}
We define the power spectrum of  tensor fluctuations as 
\begin{equation}
    \left< h_{\bm{k}}^{(\sigma)} h_{\bm{k}'}^{(\sigma)}\right> = \frac{2\pi^{2}}{k^{3}}P^{(\sigma)}(k)
                               \delta^{(3)}(\bm{k}+\bm{k}') \ .
\end{equation}
Let us divide  the tensor fluctuations into the two parts. 
The one comes from vacuum fluctuations and the other comes from the gauge field.  
Since they are uncorrelated to each other, we can write the tensor power spectrum as  the sum 
\begin{equation}
    P^{(\sigma)}(k) = P^{(\sigma)}_{v}(k)+P^{(\sigma)}_{s}(k) \label{power}\ . 
\end{equation}
From  Eqs.(\ref{hmode})$\sim$(\ref{power}),   we can deduce 
\begin{eqnarray}
    P^{(\sigma)}_{s}(k) = \frac{k^{3}}{\pi^{2}M_{pl}^{4}a^{2}} 
                &&\ \int \frac{d^{3}p}{(2\pi)^3} \left( 1+(\hat{k}\cdot\hat{p})^{2} \right) 
                 \left( 1+(\hat{k} \cdot \widehat{\bm{k}-\bm{p}})^{2} \right)  \nonumber \\
      &&  \times\left| \int d\tau' a^3 (\tau' )  G_{k}(\tau,\tau')\mathcal{B}_{p}(\tau')
      \mathcal{B}_{|\bm{k}-\bm{p}|}(\tau') \right|^{2}\ ,
      \label{power5}
\end{eqnarray}
where we ignored the subdominant contribution of the electric fields and used 
an identity 
$\left( \vec{e}^{(\alpha)}(\hat{k})\cdot \vec{e}^{(\beta)}(\hat{k}') \right)^{2}
=\frac{1}{4}\left( 1-\alpha\beta (\hat{k}\cdot\hat{k}') \right)^2$. 
Let us define the Green's function $G_{k}(\tau,\tau')$ for Eq.(\ref{hmode}). 
Substituting the scale factor in Eq.(\ref{scasol}) into 
the homogeneous part of Eq.(\ref{hmode}), we obtain 
\begin{equation}
    v_{k}''(\tau)+\left( k^{2}+\frac{2(\lambda^{2}-4)}{(\lambda^{2}-2)^{2}\tau^{2}} \right) v_{k}(\tau) 
                   = 0 \label{homog}\ .
\end{equation}
In the fast roll limit ($\lambda\rightarrow \infty$), the Green's function obtained from Eq.(\ref{homog}) becomes
\begin{eqnarray}
    G_{k}(\tau,\tau') &\equiv& \frac{\cos(k\tau)\sin(k\tau')-\sin(k\tau)\cos(k\tau')}{k} \nonumber\\ 
                      &\simeq& \tau'  \qquad 
               \left(\ \left| k\tau \right|, \ \left| k\tau' \right| \ll 1 \ \right)\label{green}\ ,
\end{eqnarray}
%
where we took the superhorizon limit since the Green's function just oscillates and does not contribute the time
integration of Eq.(\ref{power5}) in the subhorizon regime.
Substituting  Eqs.(\ref{mag}) and (\ref{green}) into  Eq.(\ref{power5}) and using the new variables $\vec{q}\equiv \frac{\vec{p}}{k}$, $\vec{q'}\equiv \frac{\vec{p}-\vec{k}}{k}$ and $z \equiv -k\tau$, we get the power spectrum 
at the end of the ekpyrotic phase as 
\begin{eqnarray}
    P_{s}(k) &=& 2\times P^{(\sigma)}_{s}(k)\nonumber \\
    &=&\frac{81}{256\pi^{5}}(\lambda^{2}-2)^{4} \left( \frac{H_{end}}{M_{pl}} \right)^{4}
      z_{end}^{\frac{4\lambda^{2}-4}{\lambda^{2}-2}} 
     \ \int d^{3}q q^{-3}q'^{-3}\left( 1+(\hat{k}\cdot\hat{q})^{2} \right) \left( 1+(\hat{k} \cdot \hat{q'})^{2} \right) 
     \nonumber\\ 
    && \hspace{5cm}
    \times\left|\int dz' z'^{-\frac{3\lambda^{2}-4}{\lambda^{2}-2}} \right|^{2} \ ,\label{se}
\end{eqnarray}
where we used the fact  that there is no polarization of gravitational waves. 
Since the gauge field becomes relevant as the source of the gravitational waves after 
the ekpyrotic phase starts, 
we consider the region  
\begin{equation}
    \left| \tau_{in} \right|\ \gtrsim \ \frac{1}{p}\ ,\ \frac{1}{\left| \vec{p}-\vec{k} \right|} \ . \label{qin} 
\end{equation}
Here,  $\tau_{in}$ is the time when the ekpyrotic phase starts. 
Multiplying it by k, we get 
\begin{equation}
    \quad \frac{1}{q_{in}} > \frac{1}{q}\ ,\ \frac{1}{q'} \quad,
\end{equation}
where $q_{in} \equiv \left|  k\tau_{in} \right|^{-1}$ represents the 
infrared cut off of the momentum integral. 
Then the momentum integral is calculated as
\begin{eqnarray}
     && \int_{q_{in}}^{1} d^{3}q q^{-3}q'^{-3}\left( 1+(\hat{k}\cdot\hat{q})^{2} \right) 
    \left( 1+(\hat{k} \cdot \hat{q'})^{2} \right) \nonumber \\
    &=&2\pi \int_{q_{in}}^{1} dq  q^{-1} \int d\theta \sin\theta
        \frac{\left( 1+\cos^{2}\theta \right) 
        \left( 1+\left(\frac{(1-q\cos\theta)}{\sqrt{1+q^{2}-2q\cos\theta}}\right)^{2}\right)}
        {\left(1+q^{2}-2q\cos\theta\right)^{3/2}} \nonumber \\
    &=& 2\pi \int_{q_{in}}^{1}\frac{16}{15}\frac{q^{4}-q^{2}-5}{q(q+1)(q-1)}dq \quad , \label{seki}
\end{eqnarray}
where we defined $\cos\theta \equiv \hat{k}\cdot \hat{q}$ and we approximately evaluated the integral in the range  $q_{in}<q<1$. 
One can see that there are two poles at $q=q_{in}\ll 1 $ and $q=1$. 
The later one is corresponding to the $q'=q_{in}\ll 1$. 
From the symmetry between $q$ and $q'$, we can evaluate the integral  (\ref{seki}) at $q=q_{in}$ by multiplying it by 2 
\begin{equation}
    \frac{64\pi}{3}\ln q_{in}^{-1}\ .
\end{equation}
On the other hand, the time integral in Eq.(\ref{se}) can be evaluated at $z'=z_{end}$ approximately. 
We therefore obtain 
\begin{equation}
    P_{s}(k) \simeq \frac{27}{16\pi^{4}} 
               \lambda^{8}
               \left( \frac{H_{end}}{M_{pl}} \right)^{4}
               \ln \left[  \frac{k}{k_{in}}  \right]\ . \label{power7}
\end{equation}
Note that we have taken the limit $\lambda \gg 1$.
There is a factor  $\left( \frac{H_{end}}{M_{pl}}\right)^{4}$ in the spectrum (\ref{power7}) because of  the nonlinear contribution of the magnetic 
fields (\ref{mag}).
One can see that sourced gravitational waves have a nearly scale invariant spectrum. 
This conclusion is different from the well-known blue-tilted spectrum in the 
ekpyrotic scenario \cite{Boyle:2003km}.  Most importantly, there appears a factor  $\lambda^{8}$ in $P_{s}(k)$. 
For example, if we set $H_{end}=10^{-5}M_{pl}\ ,\ \lambda=-17$ 
to produce the observed magnetic field, the amplitude of the power spectrum is about $10^{-11}$. 
This is comparable with the gravitational waves  in the inflationary universe $\sim \left( \frac{H_{end}}{\pi M_{pl}} \right)^{2}$. 
Therefore, we can not discriminate between inflation and the ekpyrotic  scenario
just by detecting   primordial gravitational waves. 
In the next section, let us calculate scalar fluctuations sourced by the gauge field 
and discuss if the ekpyrotic model with a gauge field is compatible with the CMB data. 
\section{Scalar Fluctuations from Magnetic Fields}
Now, we  calculate scalar fluctuations sourced by the gauge field in the present scenario. 
Too much production of scalar fluctuations implies incompatibility with the CMB  data. 
Fortunately, we will soon see that the sourced scalar fluctuations are smaller than the tensor fluctuations.
Let us show it by repeating the same procedure we used in the previous section. 
As is discussed in \cite{barnaby}, the equation for linear perturbation of 
the scalar field $\delta \phi$ in the flat slicing gauge is given by 
\begin{equation}
     s''-
       \left( \nabla^{2}+\frac{z''}{z}\right) s 
\simeq
         -a^{3}
          \left( \frac{f_{,\phi}}{f}+ \frac{\phi'}{4M_{pl}^{2}\mathcal{H}}  \right)\vec{B}^{2}\ , \label{liphi}
\end{equation}
where we ignored the subdominant contributions of the electric fields and 
used $s=a\delta \phi$, $z\equiv \frac{a\phi'}{\mathcal{H}}$. 
From Eqs.(\ref{scasol}) and (\ref{cup}), we obtain
\begin{equation}
    \frac{\phi'}{\mathcal{H}} = -M_{pl}\lambda \  ,\quad \frac{f_{,\phi}}{f}=\frac{\rho}{M_{pl}}\ . \label{fastcon}
\end{equation}
Since we are considering the scale invariant magnetic fields as the source of the scalar fluctuations, 
$\rho$ satisfies Eq.(\ref{rho}). 
Hence, working in Fourier space, 
we can rewrite Eq.(\ref{liphi}) as 
\begin{equation}
    s_{\bm{k}}''(\tau)+\left( k^{2}-\frac{a''}{a} \right) s_{\bm{k}}(\tau) 
                   \simeq S(\tau,\bm{k}) \label{smode}\ ,
\end{equation}
where the source term is defined by 
\begin{equation}
    S(\tau,\bm{k}) = -\frac{3a^{3}}{4M_{pl}^{2}} \frac{\lambda^{2}-8/3}{\lambda} \int \frac{d^{3}p}{(2\pi)^{3/2}} 
           \left( \hat{B}_{i,\bm{p}}\hat{B}_{i,\bm{k}-\bm{p}}\right)
                             \ .
\end{equation}
We define the power spectrum of scalar fluctuations as 
\begin{equation}
    \left< \left(\frac{\mathcal{H}}{\phi'}\right)\delta\phi_{\bm{k}} 
          \left(\frac{\mathcal{H}}{\phi'}\right)\delta\phi_{\bm{k'}}\right> 
          = \frac{2\pi^{2}}{k^{3}}\mathcal{P}(k)\delta^{(3)}(\bm{k}+\bm{k}') \ .  \label{sp}
\end{equation}
One can see that the fast roll condition suppresses the scalar power spectrum by the factor 
$\left( \frac{\mathcal{H}}{\phi'} \right)^{2}=\frac{1}{M_{pl}^{2}\lambda^{2}}$\ . 
The power spectrum can be divided into two parts like the tensor power spectrum as 
\begin{equation}
    \mathcal{P}(k) = \mathcal{P}_{v}(k)+\mathcal{P}_{s}(k) \label{spower}\ . 
\end{equation}
From  Eqs.(\ref{smode})$\sim$(\ref{spower}), we can deduce 
\begin{eqnarray}
    \mathcal{P}_{s}(k) = \frac{9k^{3}}{16\pi^{2}M_{pl}^{4}a^{2}} \frac{(\lambda^{2}-8/3)^{2}}{\lambda^{4}}
                &&\ \int \frac{d^{3}p}{(2\pi)^3} \left( 1+(\hat{p}\cdot\widehat{\bm{k}-\bm{p}})^{2} \right) 
                   \nonumber \\
      &&  \times\left| \int d\tau' a^3 (\tau' )  G_{k}(\tau,\tau')\mathcal{B}_{p}(\tau')
      \mathcal{B}_{|\bm{k}-\bm{p}|}(\tau') \right|^{2}\ ,
      \label{power6}
\end{eqnarray}
where the Green's function $G_{k}(\tau,\tau')$ for Eq.(\ref{smode}) is the same as that for Eq.(\ref{green}). 
Substituting  Eqs.(\ref{mag}) and (\ref{green}) into  Eq.(\ref{power6}) and using the variables $\vec{q}\equiv \frac{\vec{p}}{k}$, $\vec{q'}\equiv \frac{\vec{p}-\vec{k}}{k}$ and $z \equiv -k\tau$, we get the scalar power spectrum 
at the end of the ekpyrotic phase as 
\begin{eqnarray}
    \mathcal{P}_{s}(k) &=&\frac{729}{8192\pi^{5}}\frac{(\lambda^{2}-2)^{4}(\lambda^{2}-8/3)^{2}}{\lambda^{4}} 
    \left( \frac{H_{end}}{M_{pl}} \right)^{4}
      z_{end}^{\frac{4\lambda^{2}-4}{\lambda^{2}-2}} 
     \ \int d^{3}q q^{-3}q'^{-3}\left( 1+(\hat{q}\cdot\hat{q'})^{2} \right) 
     \nonumber\\ 
    && \hspace{7cm}
    \times\left|\int dz' z'^{-\frac{3\lambda^{2}-4}{\lambda^{2}-2}} \right|^{2} \ .\label{se2}
\end{eqnarray}
The momentum integral is carried out as 
\begin{eqnarray}
     && \int_{q_{in}}^{1} d^{3}q q^{-3}q'^{-3}\left( 1+(\hat{q}\cdot\hat{q'})^{2} \right) \nonumber \\
    &=&2\pi \int_{q_{in}}^{1} dq  q^{-1} \int d\theta \sin\theta
        \frac{ \left( 1+\left(\frac{(q-\cos\theta)}{\sqrt{1+q^{2}-2q\cos\theta}}\right)^{2}\right)}
        {\left(1+q^{2}-2q\cos\theta\right)^{3/2}} \nonumber \\
    &=& -2\pi \int_{q_{in}}^{1}\frac{8}{3}\frac{1}{q(q+1)(q-1)}dq \quad , \label{seki2}
\end{eqnarray}
where $\cos\theta \equiv \hat{k}\cdot \hat{q}$ and we approximately evaluated the integral in the range  $q_{in}<q<1$. 
From the symmetry between $q$ and $q'$, we can calculate Eq.(\ref{seki2}) at $q=q_{in}$ by multiplying it by 2 as is done for 
tensor fluctuations. The result reads
\begin{equation}
    \frac{32\pi}{3}\ln q_{in}^{-1}\ .
\end{equation}
The time integral is same as the case of tensor fluctuations and we can obtain the scalar power spectrum 
sourced by the scale invariant magnetic fields as
\begin{equation}
    \mathcal{P}_{s}(k) \simeq \frac{243}{1024\pi^{4}} 
               \lambda^{8}
               \left( \frac{H_{end}}{M_{pl}} \right)^{4}
               \ln \left[ \frac{k}{k_{in}}  \right]\ .  \label{power8}
\end{equation}
From Eqs.(\ref{power7}) and (\ref{power8}), the tensor to scalar ratio $r_{source}$ is given by
\begin{equation}
    r_{source} \simeq 7  \label{ratio}
                 \ .
\end{equation}
This result is different from that in the inflationary universe, where the scalar fluctuations 
are  enhanced by the inverse square of a slow roll parameter  \cite{barnaby}. 
Taking a look at terms in the parenthesis of the right-hand side of Eq.(\ref{liphi}), 
we see that it gives rise to a factor $\lambda^{2}$ in the scalar power spectrum in contrast  to the  case of tensor fluctuations. 
On the other hand, from Eq.(\ref{sp}), we see the scalar power spectrum is suppressed by $\frac{1}{\lambda^{2}}$ 
in contrast to the case of tensor fluctuations. These two factors have been canceled out. 
The numerical value (\ref{ratio})  comes from accumulation of several factors such as the polarization degrees of freedom. 
Since the tensor to scalar ratio becomes larger  than unity, 
we can say that the scalar fluctuations sourced by the scale invariant magnetic field are negligible 
in the ekpyrotic scenario due to the fast roll condition.
\section{Conclusion}
We studied the role of the gauge kinetic function in the ekpyrotic scenario and showed that 
abundant gravitational waves sourced by the gauge field can be produced. 
As a demonstration, we first showed that scale invariant magnetic fields can be produced in the ekpyrotic phase. 
It turned out that the magnetic fields induce nearly scale invariant gravitational waves (slightly blue) and 
the amplitude could be comparable with that of the inflationary universe. 
It turned out that it is difficult to  disprove the ekpyrotic scenario  by  detecting  primordial gravitational waves. 
In order to distinguish both scenarios, it is necessary to look at the details 
of the spectrum such as the tilt of the spectrum. 
Observing the distinction of higher order scalar perturbations is also important \cite{Chen:2016cbe}. 
We should mention that the idea of finding an ekpyrotic model with observable gravitational waves 
on CMB scales using sourced fluctuations was put forward for the first time in \cite{Ben-Dayan:2016iks} by investigating a different model with explicit parity violation. Our model has no explicit parity violation. Moreover, we also showed that the scalar fluctuations induced by the magnetic field are smaller than the sourced gravitational waves. 
Generally, as far as the fast roll condition is satisfied, the  tensor to scalar ratio becomes 
more than unity in any ekpyrotic models with the gauge kinetic function. 
Therefore, our scenario would be compatible with the CMB data provided that  nearly scale invariant scalar fluctuations 
are produced in a standard way with an additional scalar field \cite{Levy:2015awa}. 

It should be noted that we must check the non-gaussianity of the primordial scalar fluctuations in the present model 
\cite{Ijjas:2014fja}.  Moreover,  we should consider a bounce process from contracting to expanding 
to connect the spectrum at the end of the ekpyrotic phase with observables.
We have not looked into this issue in this paper since the mechanism is model dependent and 
the detailed analysis is beyond the scope of this paper \cite{Brandenberger:2016vhg}. 
However, actually, although we fixed the parameters such as $\rho, \lambda, H_{end}$ for simplicity in this paper, 
we can tune these parameters in our scenario so that our conclusion becomes valid for any ekpyrotic bouncing models. 
Therefore, our conclusion is robust. 
\acknowledgements
This work was in part supported by MEXT KAKENHI Grant Number 15H05895. 

\end{document}